\documentstyle[twocolumn,aps,psfig]{revtex}

\begin{document}
\draft

\twocolumn[\hsize\textwidth\columnwidth\hsize\csname @twocolumnfalse\endcsname

\title{
Field-induced long-range order in the $S=1$ antiferromagnetic chain
}

\author{T\^oru Sakai}
\address{
Faculty of Science, Himeji Institute of Technology, 3-2-1 Kouto, Kamigori,
Ako-gun, Hyogo 678-1297, Japan
}

\date{July 2000}
\maketitle 

\begin{abstract}
The quasi-one dimensional $S=1$ antiferromagnet in magnetic field $H$ is 
investigated with the exact diagonalization of finite chains and the 
mean field approximation for the interchain interaction. 
In the presence of the single-ion anisotropy $D$, 
the full phase diagram in the $HT$ plane 
is presented for $H\parallel D$ and $H\perp D$. 
The shape of the field-induced long-range ordered phase is revealed 
to be quite different between the two cases, as observed in the recent
experiment of NDMAP. 
The estimated ratio of the interchain and intrachain couplings 
of NDMAP  
($J'/J \sim 10^{-3}$) is consistent with the neutron scattering 
measurement. 
\end{abstract}

\vskip2pc]
\narrowtext

%
%

Since Haldane conjectured\cite{haldane} 
the presence of the spin excitation gap 
in the one-dimensional integer-$S$ Heisenberg antiferromagnets, 
the $S=1$ antiferromagnetic chain has attracted a lot of interest. 
Many theoretical and numerical works indicated the disordered 
ground state of the system. 
The mean field approximation for the interchain interaction 
combined with the exact diagonalization of finite chains\cite{sakai1} 
revealed that such a disordered ground state should be realized 
even in the quasi-one-dimensional $S=1$ antiferromagnet 
if the ratio of the interchain and intrachain couplings holds 
$zJ'/J \lesssim 0.05$ where $z$ is the number of adjacent chains. 
The criterion was satisfied for the material 
Ni(C$_2$H$_8$N$_2$)$_2$NO$_2$(ClO$_4$)\cite{nenp0},
abbreviated NENP. 
In fact the Haldane gap had already been observed in the 
temperature dependence of the magnetic susceptibility\cite{nenp0} and 
the high-field magnetization measurements of NENP.\cite{nenp1,nenp2} 

The effective field theory\cite{affleck} 
and the size scaling analysis with the 
finite chain calculation\cite{sakai2,sakai3} 
indicated that the field-induced transition 
occurs from the disordered ground state to the gapless
Tomonaga-Luttinger liquid phase at some critical external field 
$H_{c1}$ where the Haldane gap vanishes. 
With the interchain interaction, the gapless phase would become 
the ordered phase. 
Thus the field-induced antiferromagnetic long-range order 
is expected to appear at $H_{c1}$ in the quasi-one-dimensional 
$S=1$ system. 
NENP, however, didn't exhibit such a field-induced spontaneous 
symmetry breaking, because the staggered moment is forced to 
appear even by a smaller external field through the alternation 
of the $g$-tensors.\cite{chiba,mitra,sakai4} 
Recently the field-induced transition to the long-range 
order was discovered in the high-field magnetization measurement 
of novel quasi-one-dimensional $S=1$ systems 
Ni(C$_5$H$_{14}$N$_2$)$_2$N$_3$(ClO$_4$)\cite{ndmaz} and 
Ni(C$_5$H$_{14}$N$_2$)$_2$N$_3$(PF$_6$)\cite{ndmap}, 
abbreviated NDMAZ and 
NDMAP, respectively.  
The magnetic susceptibility measurement indicated the presence of 
the easy-plane single-ion anisotropy $D$ for both systems and the value was 
reported as $D \sim 0.3 J$ for NDMAP. 
The experimental phase diagram of both systems in the $HT$ plane 
exhibited a quite different shape of the ordered phase 
between the two cases when the external field $H$ is (a)parallel 
and (b)perpendicular to the principal axis of $D$.  
This is probably because the universality of the symmetry breaking 
is different between the two cases. 
Thus it would be interesting to present the theoretical $HT$ phase 
diagram. 
In this paper, 
using the exact diagonalization of finite chains and the mean field 
approximation for the interchain interaction, 
the phase diagrams for $H\parallel D$ and $H\perp D$ 
are presented, 
to compare with the experimental ones and give some suggestions 
for future experiments.  

In order to investigate the magnetization process of the 
quasi-one-dimensional $S=1$ antiferromagnet with the planar 
single-ion anisotropy, we consider the Hamiltonian,
\begin{eqnarray}
\label{ham}
{\cal H}& =& J\sum _j\sum_i {\bf S}_{j,i} \cdot {\bf S}_{j,i+1}  
+ J'\sum _{\langle j,j'\rangle}\sum_i {\bf S}_{j,i} \cdot {\bf S}_{j',i} 
\nonumber \\ 
&+& D \sum_j \sum_i (S_{j,i}^z)^2 -H\sum_j\sum_i S_{j,i}^{\alpha},
\end{eqnarray}
where $J$ and $J'$ are the intrachain and interchain couplings,
respectively, $j$ is the index of a chain and $\sum _{\langle j,j'\rangle}$
means the sum over all the pairs of nearest-neighbor chains. 
We set $J=1$ and use the units such as $g\mu _B=1$ throughout the paper.
Only the $XY$-like anisotropy $D>0$ is investigated, to compare with the
experimental results of NDMAP. 
To consider the two cases (a)$H\parallel D$ and (b)$H\perp D$, 
we set $\alpha=z$ and $y$, respectively. 
When the antiferromagnetic order is possibly induced by the external 
field $H$, $xy$-plane is an easy plane for (a) while $x$-axis is an easy
axis for (b). 
Thus the universality class of the symmetry breaking 
should be different between the two cases; the former corresponds 
to the $XY$ model ($U(1)$) while the latter the Ising one ($Z_2$). 

In the magnetization process at low temperatures two phase transitions
are expected to occur at critical fields $H_{c1}(T)$ and $H_{c2}(T)$. 
At $T=0$ the Haldane gap vanishes at $H_{c1}(0)$ and the magnetization
is saturated at $H_{c2}(0)$. 
The antiferromagnetic long-range order in the vertical direction to $H$
should appear between $H_{c1}(T)$ and $H_{c2}(T)$. 
Based on the mean field approximation for the interchain interaction, 
the critical field $H_{c1}(T)$, as well as $H_{c2}(T)$ 
is determined by the solution $H$ of the equation  
\begin{eqnarray}
\chi _{\rm st (1D)}^{\rho}(T,H) ={1\over {zJ'}},
\label{staggered}
\end{eqnarray}
where $\chi _{\rm st (1D)}^{\rho}(T,H)$ is the staggered susceptibility
of the one-dimensional system ($J'=0$) and $\rho$ (=$x$, $y$ or $z$) 
specifies the easy component of the antiferromagnetic order. 
To consider the two cases (a) and (b) we set 
$(\alpha, \rho)=(z,x)$ and $(y,x)$, respectively. 
Using the numerical diagonalization analysis, 
we obtain the all the eigenvalues and eigenvectors of finite chains 
and calculate $\chi _{\rm st (1D)}^{\rho}(T,H)$. 
Since the rotational symmetry doesn't remain in any directions for 
$H\perp D$, the available chain length is at most $L=8$ under the periodic
boundary condition because of limitation of the memory size.  
At finite temperature, however, 
the correlation length is much smaller than that at $T=0$ 
($\xi \sim$5 to 6 lattice constants for $H=D=0$). 
Thus even the result for $L=8$ is expected to indicate some qualitative
features of the bulk system.  
In fact the small oscillation in the $H$ dependence of 
$\chi _{\rm st (1D)}^{\rho}(T,H)$ due to the finite 
size effect only appears at $T \lesssim 0.1$ for $L=8$. 
Based on obtained $H_{c1}(T)$ and $H_{c2}(T)$ from the calculation 
for $L=8$, we discuss on the properties of the phase diagram in the  
$HT$-plane in the following. 

%
%
\begin{figure}[htb]
\begin{center}
\mbox{\psfig{figure=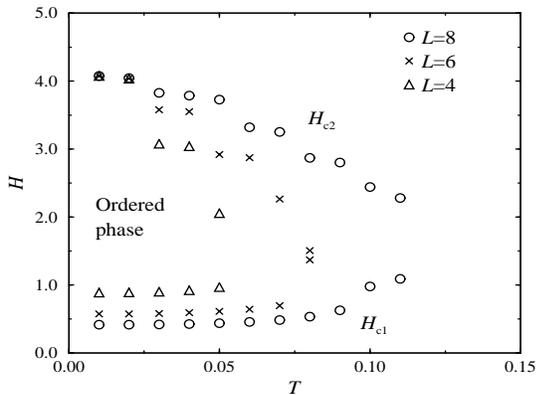,width=8cm,height=6cm,angle=-90}}
\end{center}
\caption{$H_{c1}(T)$ and $H_{c2}(T)$ plotted versus $T$ for 
$L=$8, 6 and 4. The finite size effect is revealed to underestimate 
the ordered phase.  
\label{fig1}
}
\end{figure}

To investigate the finite size effect, 
we show in Fig. \ref{fig1} 
$H_{c1}(T)$ and $H_{c2}(T)$ of the isotropic system ($D=0$) with 
$zJ'=0.05$ determined by the equation 
(\ref{staggered}) for $L=$8, 6 and 4 in the $HT$-plane. 
The lower and upper boundaries of the antiferromagnetic ordered phase 
are given by $H_{c1}(T)$ and $H_{c2}(T)$, respectively. 
Those lines also give the $H$ dependence of 
the N\'eel temperature $T_{\rm N}$.  
Fig. \ref{fig1} indicates that 
estimated $T_{\rm N}$ is higher for larger $L$, 
which suggests that the finite-chain analysis tends to 
underestimate the ordered phase. 
On the other hand, 
the mean field approximation generally overestimates it.  
Thus on the present analysis the finite size effect and 
the fluctuation around the mean field are expected to 
suppress the errors due to each other.

%
%
\begin{figure}[htb]
\begin{center}
\mbox{\psfig{figure=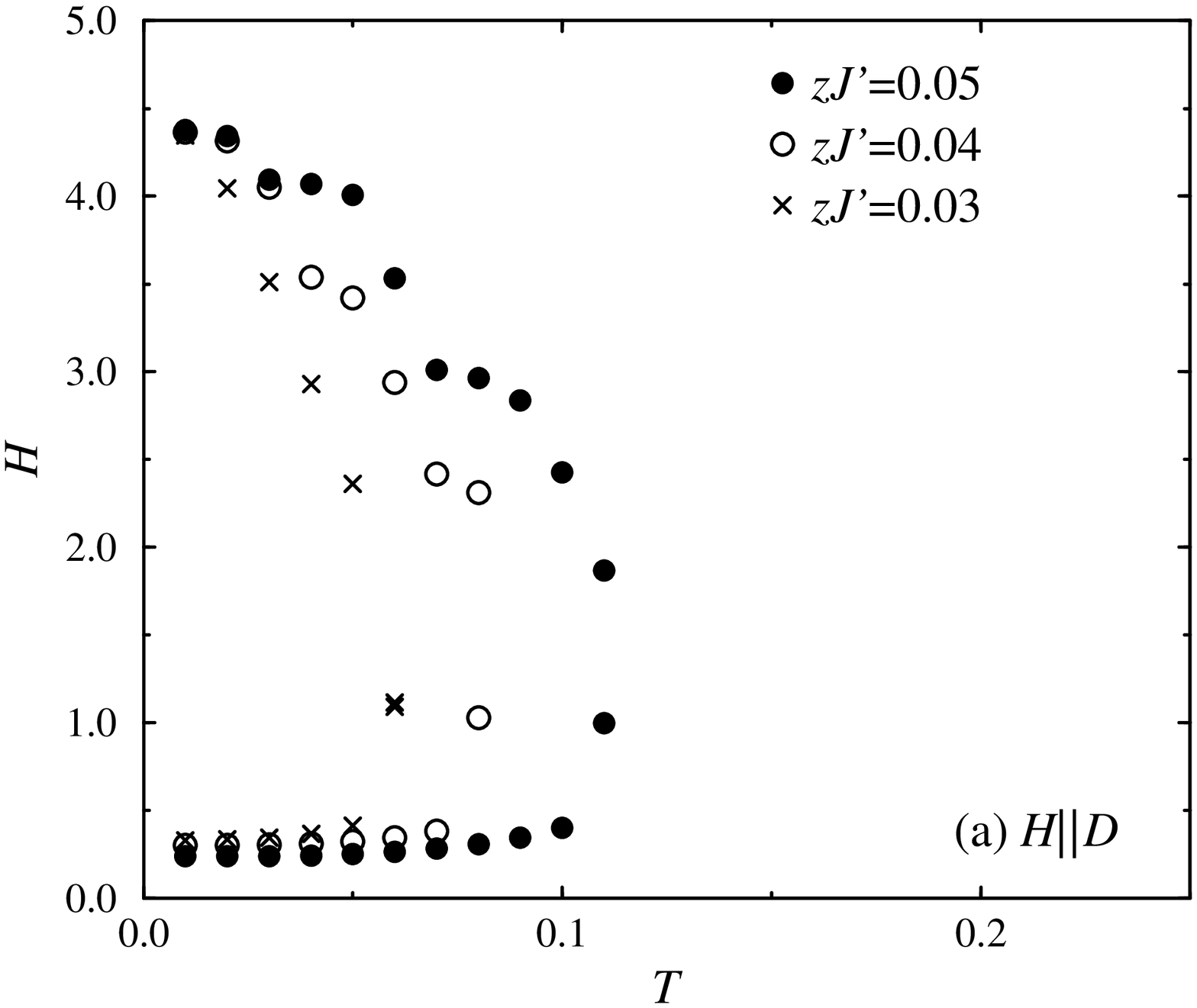,width=8cm,height=6cm,angle=-90}}
\mbox{\psfig{figure=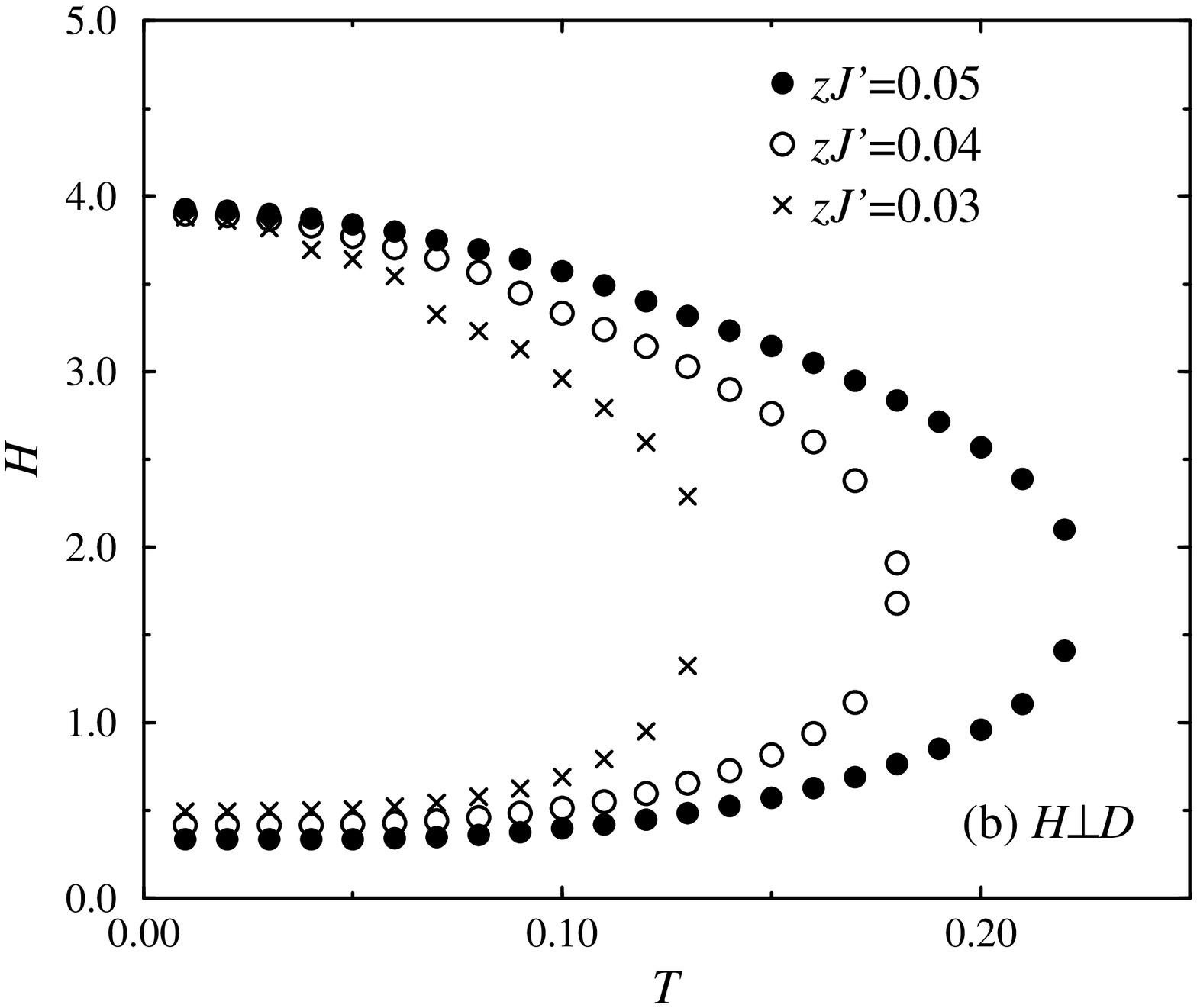,width=8cm,height=6cm,angle=-90}}
\end{center}
\caption{
Phase diagrams in the $HT$-plane for (a)$H\parallel D$ and 
(b)$H\perp D$ with $D=0.3$. 
\label{fig2}
}
\end{figure}

In the following analysis we put $D=0.3$, 
realistic for NDMAP. 
The phase diagrams in the $HT$-plane obtained from 
$\chi _{\rm st (1D)}^{\rho}(T,H)$ for $L=8$ are shown in 
Figs. \ref{fig2}(a) and (b), for $H\parallel D$ and $H\perp D$, 
respectively. 
The phase boundaries for $zJ'=$0.05, 0.04 and 0.03 are given 
in each case. 
Although $H_{c2}(T)$ exhibits an oscillation due to finite size 
effect, Fig. \ref{fig2}(a) clarifies an interesting feature; 
$H_{c1}(T)$ and $H_{c2}(T)$ are quite asymmetric for $H \parallel D$. 
In contrast, 
$H$ dependence of $T_{\rm N}$ is close to a symmetric bell shape 
for $H\perp D$. 
According to our calculation for other values of $D$, 
the difference in the shape of $T_{\rm N}$ becomes clearer 
for larger $D$.  
With increasing $D$, 
$T_{\rm N}$ becomes more asymmetric for $H\parallel D$,
while closer to a completely symmetric bell shape for $H\perp D$. 
The asymmetric behavior of $T_{\rm N}$ for $H\parallel D$ 
is consistent with the 
$H$ dependence of the critical exponent $\eta$ in the transverse 
spin correlation function 
$\langle S_0^xS_r^x\rangle \sim (-1)^r r^{-\eta}$ of the 
isotropic $S=1$ chain at $T=0$.    
In the ground-state magnetization process of the system 
the antiferromagnetic spin correlation measured by $\eta$ 
rapidly grows from $H_{c1}$ to the maximum while  
slowly decays toward $H_{c2}$. 
Thus such an asymmetric behavior is expected to be characteristic 
of the system with $U(1)$ symmetry. 
On the other hand, 
the case (b) with only $Z_2$ symmetry is supposed to be close to 
a classical spin system where the spin wave excitation spectrum 
around $m\sim 1$ is similar to that around $m\sim 0$. 

\begin{figure}[htb]
\begin{center}
\mbox{\psfig{figure=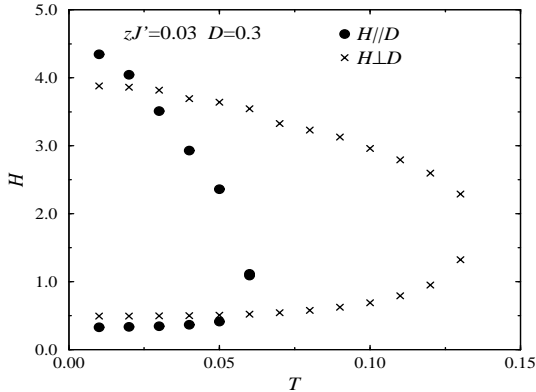,width=8cm,height=6cm,angle=-90}}
\end{center}
\caption{
Phase diagrams for $H\parallel D$ and $H\perp D$ with $zJ'=0.03$ 
and $D=0.3$. 
\label{fig3}
}
\end{figure}

The phase diagrams for $H\parallel D$ and $H\perp D$ ($zJ'=0.03$) 
are shown together in Fig. \ref{fig3}. 
They exhibit several characteristic behaviors observed in the 
experiments of NDMAP and NDMAZ; 
(i)At lower temperatures $H_{c1}(T)$ is smaller 
for $H\parallel D$. 
(ii)At intermediate $H$ $T_{\rm N}$ is higher for $H\perp D$. 
(iii)The two $T_{\rm N}$ curves have two intersections, 
although the upper one has not been observed because such a 
strong magnetic field cannot be realized. 
(i) and (ii) imply that the ordered state resulting from 
the $U(1)$ ($Z_2$) symmetry breaking is less (more) sensitive 
to the external magnetic field, while less (more) stable 
against the thermal fluctuation. 
Our calculation for various values of $D$ indicates that 
the intersection of two $T_{\rm N}$ curves appears only 
for $D>0$. 
Thus the observed intersection is also an evidence of the 
positive $D$ for NDMAP and NDMAZ. 
It is consistent with the sign of $D$ determined by the 
susceptibility measurements.\cite{ndmap,ndmaz} 
 
%
%

%
%
\begin{figure}[htb]
\begin{center}
\mbox{\psfig{figure=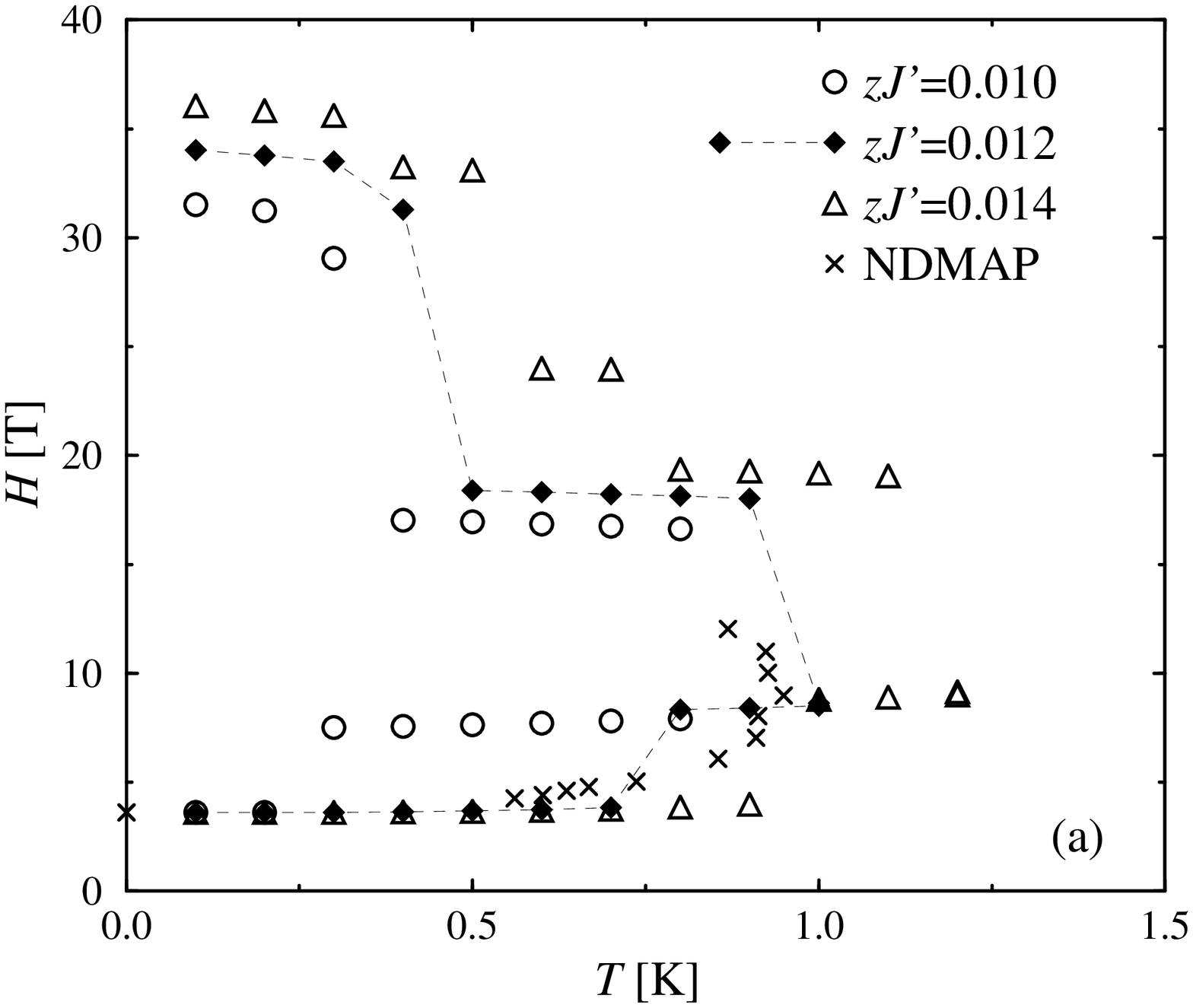,width=8cm,height=6cm,angle=-90}}
\mbox{\psfig{figure=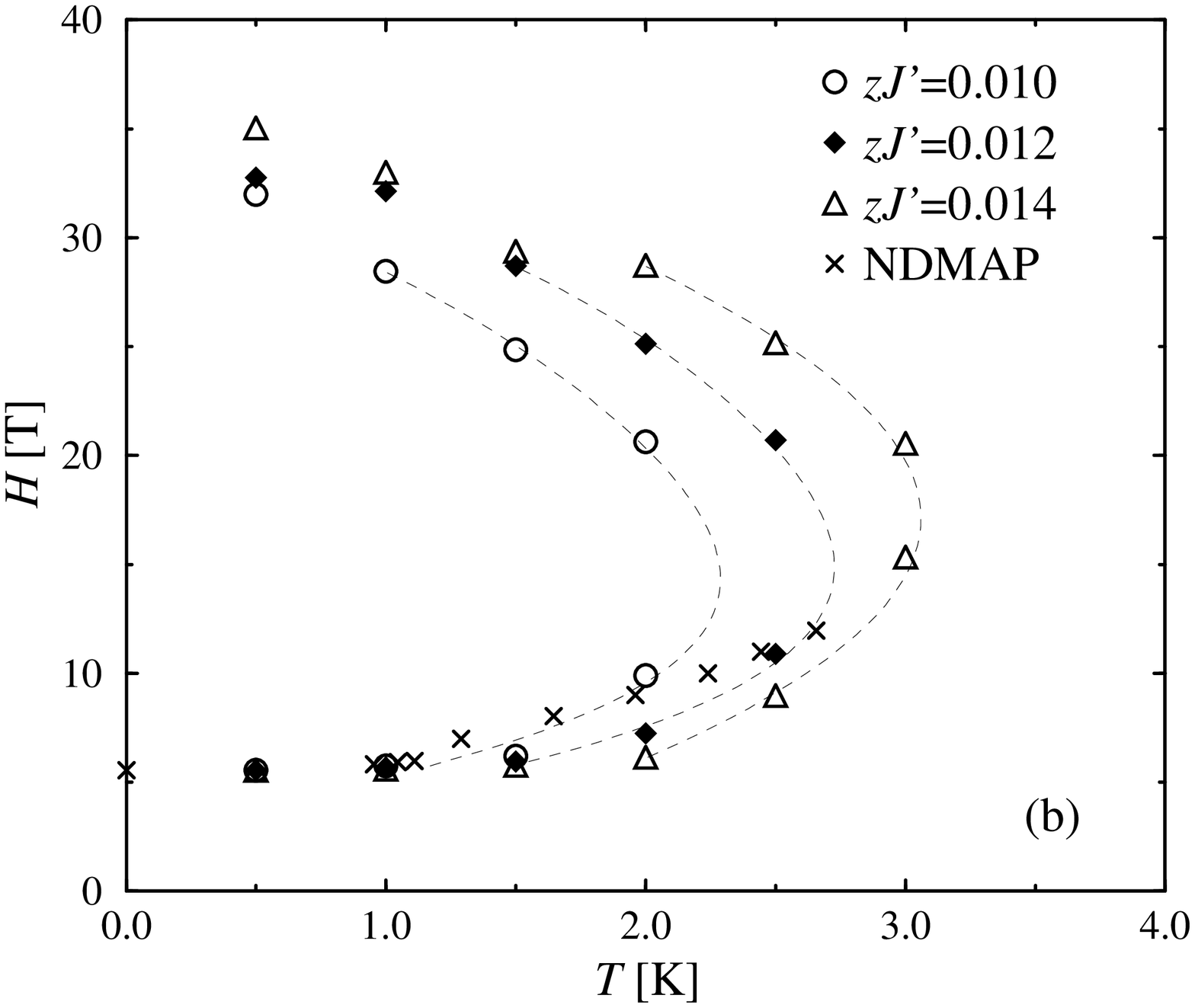,width=8cm,height=6cm,angle=-90}}
\end{center}
\caption{
Calculated phase diagrams for $zJ'$=0.010, 0.012 and 0.014, 
shown with the experimental results of NDMAP, in the cases of 
(a)$H\parallel c$-axis and (b)$H\perp c$-axis. 
Dashed curves are guides for the eyes. 
\label{fig4}
}
\end{figure}

The qualitative consistency of the present results with the 
experiments of NDMAP encourages us to estimate the strength of 
the interchain coupling $J'$ of NDMAP by fitting the calculated 
phase boundaries to the measured one. 
The calculated phase boundaries ($H_{c1}(T)$ and $H_{c2}(T)$) 
for $zJ'=$0.010, 0.012 and 0.014 are shown together with the 
experimental results of NDMAP in Figs. \ref{fig4}
(a)$H\parallel D$ and (b)$H\perp D$, where dashed curves are 
guides for the eyes. 
We set $D=0.3$. 
The temperature is normalized as $J$=30K and the magnetic field 
is normalized such that $H_{c1}(0)$ coincides with the experimental 
results.  
In Fig. \ref{fig4} (a) it is difficult to fit the calculated 
$T_{\rm N}$ curve to the experimental one, because the finite 
size effect is still large. 
Thus we determine $zJ'$ by the agreement of the maximum value 
of $T_{\rm N}$ as $zJ'=0.012 \pm 0.002$. 
On the other hand, 
the fit of the calculated $T_{\rm N}$ to the measured one 
is much better for $H\perp D$ in Fig. \ref{fig4}(b). 
It leads to the same conclusion $zJ'=0.012 \pm 0.002$.
However, the error of the estimation might be much larger, 
because we neglect the finite size correction and the fluctuation 
around the mean field of the interchain couplings. 
Since the number of the adjacent chains is usually more than 2, 
the result $zJ'/J \sim 0.012$ leads to $J'/J \sim 10^{-3}$. 
It is consistent with the most dominant interchain coupling 
estimated from the recent neutron scattering experiment.\cite{neutron} 
Using the same mean field approximation with the transfer matrix 
calculation of the chain, a little smaller value of $zJ'/J$ was 
estimated.\cite{harada} 
It is a reasonable difference, because the size correction of the 
present analysis would shift the result to a smaller value.
However, it is difficult to determine which result is better, 
because some analyses beyond the mean field approximation 
should modify it into a larger value. 
Fig. \ref{fig4} (b) suggests that the maximum of $T_{\rm N}$ 
of NDMAP could be detected with a little larger external field 
even for $H\perp D$.  
It would also enable us to perform a more precise estimation of $J'$ 
and confirm the difference in the shape of $T_{\rm N}$ between 
$H\parallel D$ and $H\perp D$. 

Recently a sign of the field-induced long-range order was also 
observed in the NMR measurement\cite{goto} of another quasi-one-dimensional 
$S=1$ system (CH$_3$)$_4$NNi(NO$_2$)$_3$, abbreviated TMNIN. 
Since $J$ is smaller ($J\sim 12$K), 
the full magnetization curve has already been obtained.\cite{tmnin} 
Thus the full phase diagram in the $HT$-plane will possibly 
be determined in the near future.  
It would clarify more detailed features of the field-induced 
long-range order.

In summary, 
based on the mean field approximation for the interchain interaction 
and numerical diagonalization of finite chains,  
we obtained the phase diagram in the $HT$-plane of the
quasi-one-dimensional $S=1$ antiferromagnet. 
The results revealed that the shape of the field-induced ordered 
phase is quite different between $H\parallel D$ and $H\perp D$. 
Fitting the calculated phase boundary to the experimental 
result lead to the estimation of the interchain coupling of 
NDMAP as $zJ'/J\sim 0.012$, which well agrees with the 
estimation by the recent neutron scattering measurement 
($J'/J\sim 10^{-3}$).

%
%
We thank T. Goto, Z. Honda, K. Katsumata and I. Harada for 
fruitful discussions. 
The computation in this work has been done using the
facilities of the Supercomputer Center, Institute for Solid
State Physics, University of Tokyo.
This research was supported in part by Grant-in-Aid 
for the Scientific Research Fund from the Ministry 
of Education, Science, Sports and Culture (11440103).

\end{document}